\documentclass[twocolumn,prl,aps,superscriptaddress,showpacs,amsmath,amssymb,floatfix]{revtex4-1}
\usepackage{color}
\usepackage{xcolor}
\usepackage{mathrsfs}
\usepackage{amsmath}
\usepackage{graphicx}% Include figure files
\usepackage{dcolumn}% Align table columns on decimal point
\usepackage{bm}% bold math

\begin{document}

\title{Unconventional Fulde-Ferrel-Larkin-Ovchinnikov states in spin-orbit coupled condensates: exact results}

\author{Qiang Han}
\affiliation{Department of Physics, Renmin University of China, Beijing, China} %
\affiliation{Department of Physics and Center of Theoretical and Computational Physics,The University of Hong Kong, Pokfulam Road, Hong Kong China}%

\author{Jia Liu}
\affiliation{Department of Physics and Center of Theoretical and Computational Physics,The University of Hong Kong, Pokfulam Road, Hong Kong China}%

\author{Dan-Bo Zhang}
\affiliation{Department of Physics, Renmin University of China, Beijing, China} %

\author{Z.~D. Wang}
\affiliation{Department of Physics and Center of Theoretical and Computational Physics,The University of Hong Kong, Pokfulam Road, Hong Kong China}%

\date{\today}

\begin{abstract}
We find that a model Hamiltonian of $s$-wave superconductors in the presence of spin-orbit interactions and
a Zeeman field is exactly solvable. Most intriguingly, based on the exact solutions, an unconventional type of  Fulde-Ferrel-Larkin-Ovchinnikov (FFLO) ground state is rigorously revealed,
in which the center-of-mass momentum of the fermion pair is proportional to the Zeeman field. We also elaborate on the drifting effect of the Zeeman field
on the spin-orbit-coupled Bose-Einstein condensate.

\end{abstract}


\maketitle

Topological superconductors are expected to be potential platforms for topological quantum computation and therefore have attracted
great interests recently in the fields of condensed matter, cold atoms, and quantum computation.
As is known, gapless edge-modes are topologically protected against any weak disorder that cannot destroy the bulk gap and break the relevant symmetries.
 It was long ago proposed that non-Abelian Majorana fermions supported by the topological superconducting phase may be realized in
the chiral $p_x + ip_y$ superconductors \cite{Read}.
Recently, a rather simple realization of topological
superconductors was theoretically suggested by inducing the effective chiral $p$-wave superconductivity on the
surfaces of topological insulators \cite{Fu} or semiconductor films \cite{Sau} in proximity to an $s$-wave superconductor. Both suggestions %the proposed topological insulator and the semiconductor film
are in the framework of mean-field theory (MFT) and involve crucially spin-orbit (SO) interactions related to Dirac fermions.
%For the above two cases, it was not until
Very recently, the
pairing Hamiltonian of Dirac fermions in two dimension (2D) was shown to be exactly solvable \cite{JLiu} and an equivalent chiral $p$-wave ground state
was rigorously confirmed beyond the mean-field approximation.

On the other hand, 3D topological superconductors \cite{Hor,Wray,Kriener} were very recently reported in the copper-doped topological insulator Bi$_2$Se$_3$.
The Fermi energy of the superconducting sample was found to be in the relativistic regime of bulk band dispersion, indicating that the electrons participating
superconductivity are also Dirac fermions \cite{Wray}. Moreover, it was revealed that the effective spin-orbit interactions or Dirac fermions can also be simulated with controllable experimental parameters using ultracold atoms in optical lattices both theoretically~\cite{zhu1,zhu2} and experimentally~\cite{lin2011},
including 3D cases even with a rather flat band~\cite{zhu3}.
Therefore, it is natural and critical to ask: (i) whether the BCS-type model of Dirac fermions is exactly solvable for 3D cases; (ii) how to solve it if the answer is yes; (iii) whether the existing important conclusions for the 2D are valid or not for the 3D; and (iv) whether an exotic quantum state of Dirac fermions can be emergent in the presence of Zeeman field.

%The study of the nature of superconductivty in this topological superconductor is timely, and the relation %and comparison to 2D superconductors of Dirac fermions are also instructive.

In the present work, we answer the above interesting and important questions timely and unambiguously.
We study the SO-coupled superfluidity in the presence of an effective Zeeman field. Conventional $s$-wave superconductors subject to a sufficient strong Zeeman field might exhibit the so-called Fulde-Ferrel-Larkin-Ovchinnikov (FFLO) state \cite{FFLO}.
However, the conventional FFLO state is rather weak, such that  concrete experimental verifications of this intriguing state are still highly awaited. While, the situation is distinctly different for pairing Dirac fermions because
the Zeeman field acts on the Dirac fermions just like an effective vector potential \cite{Yokoyama,JLiu},
and consequently a new type of unconventional FFLO state may be emergent, with the center-of-mass momentum of Cooper pairs being proportional to the Zeeman field.
Here we solve the FFLO ground state exactly, and elaborate that it is stable at least for a special case and may be observed more easily in experiments.

\begin{figure}[ht]
\begin{tabular}{cc}
\includegraphics[width=3cm]{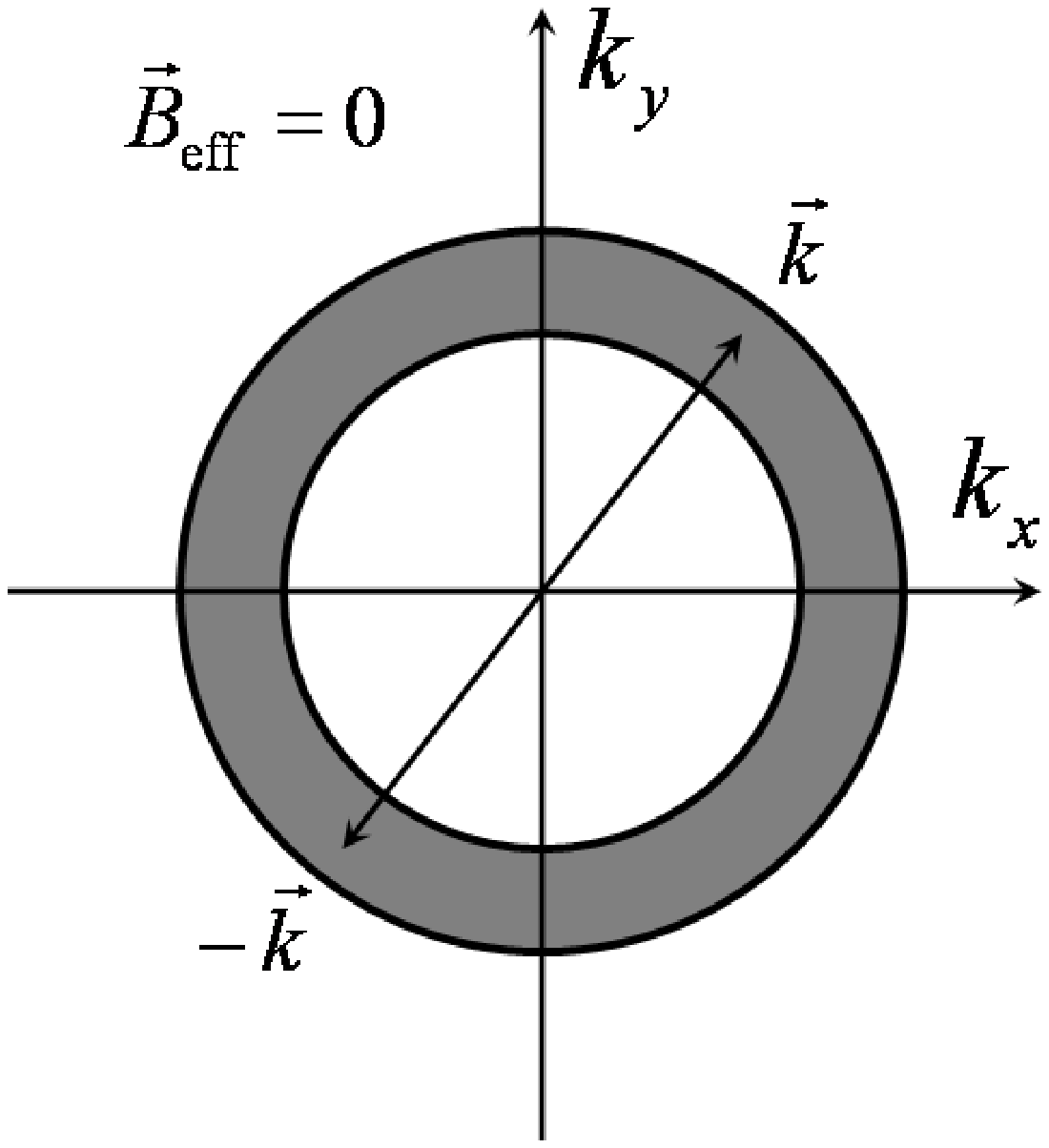} & \includegraphics[width=3cm]{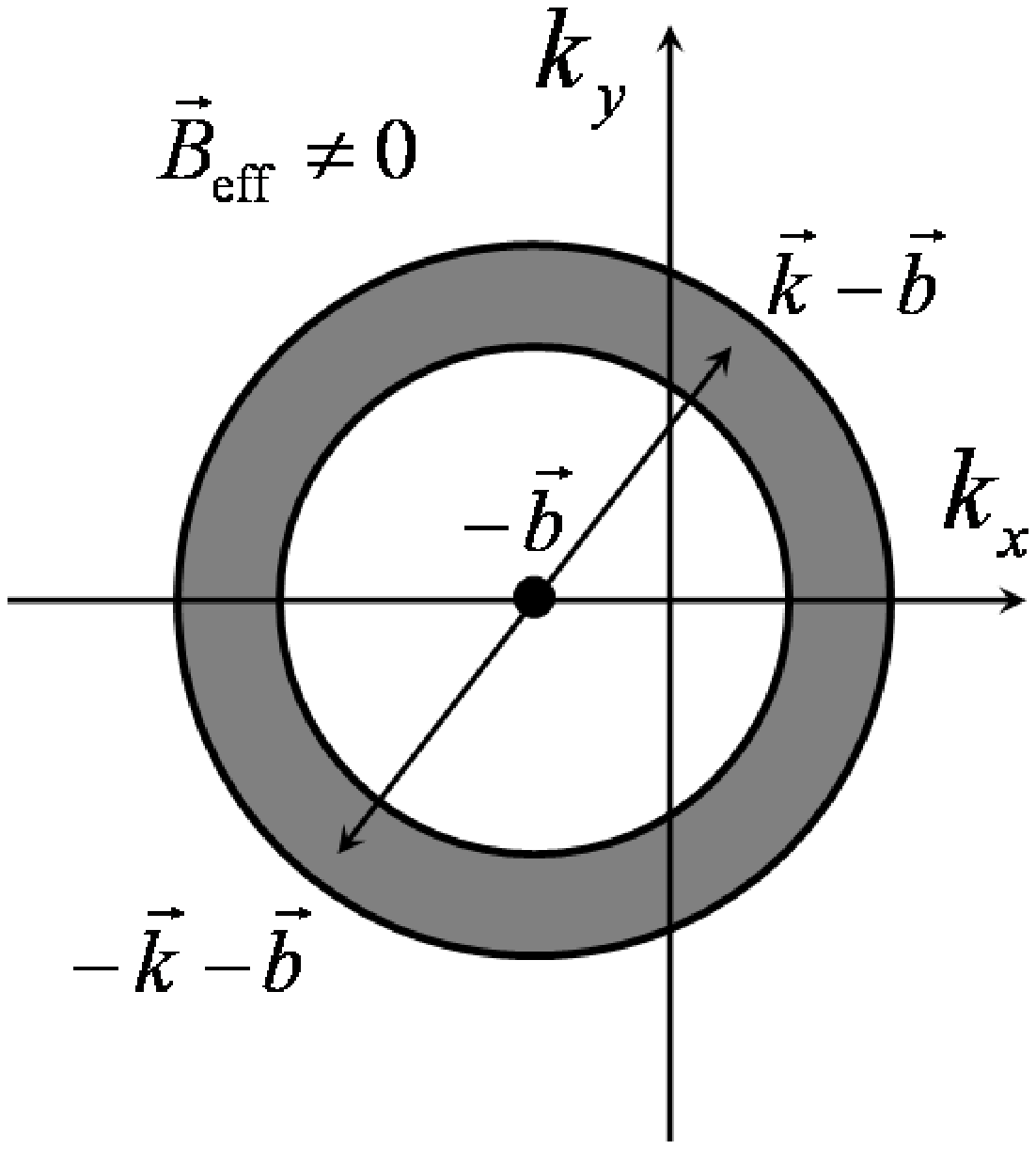}
\end{tabular}
\caption{Illustration of the evolution of the Fermi surface and energy shell with $\mathbf{B}$ for the case where
$\varepsilon_\mathbf{k}$ is constant or weakly $\mathbf{k}$ dependent. The shaded areas denote the energy shells around the Fermi surfaces.
}
\label{demo}
\end{figure}

We consider a fermionic pairing Hamiltonian with spin-orbit interactions in a 3D system
in the presence of a general effective {\it Zeeman} field $\mathbf{B}=(B_x,B_y,B_z)$ coupled with the spin degrees of freedom. The total Hamiltonian reads
\begin{equation}
\hat{H} =\hat{H}_{0}+ \hat{H}_\text{int}, \label{ham}
\end{equation}
with
\begin{equation}
\begin{aligned}
& \hat{H}_{0} = \sum_{{\bf{k}}}(c_{{\bf{k}}\uparrow}^{\dag},c_{{\bf{k}}\downarrow}^{\dag})
                (\varepsilon_{{\bf{k}}}+[\alpha\mathbf{k}+\lambda\mathbf{B}]\cdot{\bm{\sigma}})(c_{{\bf{k}}\uparrow},c_{{\bf{k}}\downarrow})^\text{T}, \\
& \hat{H}_\text{int}= \sum_{{\bf{k}},\bf{k}',\bf{q}} V_0(\mathbf{k},\mathbf{k}^\prime) c_{{\bf{k+q}}\uparrow}^{\dag}c_{{\bf{-k+q}}\downarrow}^{\dag}c_{{\bf{-k'+q}}\downarrow}c_{{\bf{k'+q}}\uparrow},
\end{aligned}
%\label{}
\end{equation}
where $\mathbf{k}=(k_x,k_y,k_z)$ is the wave-vector and ${\bm{\sigma}}=(\sigma_x,\sigma_y,\sigma_z)$ with $\sigma_{x,y,z}$ the pauli matrices.
$\varepsilon_\mathbf{k}=\varepsilon_\mathbf{-k}$ is the
spin-independent single-particle energy. $\alpha$ and $\lambda$ parametrize the SO and Zeeman coupling strengthes, respectively.  $V_0(\mathbf{k},\mathbf{k}^\prime)=V_0<0$. For physical relevance, the Zeeman field $\mathbf{B}$ may correspond to the laser-induced effective Zeeman field \cite{zhu2,Liu} in ultracold fermionic systems (or the magnetic field applied within the plane of 2D electron pairing systems). The single-particle
part of the Hamiltonian, i.e. $\hat{H}_0$, can be diagonalized by the unitary transformation $(c_{{\bf{k}}\uparrow},c_{{\bf{k}}\downarrow})^\text{T}
=U_\mathbf{k+b} (a_{\mathbf{k}+},a_{\mathbf{k}-})^\text{T}$ with $\mathbf{b}=\lambda\alpha^{-1}\mathbf{B}$ and
\begin{equation}
U_\mathbf{k} =
\left(
        \begin{array}{cc}
        \cos\theta_\mathbf{k} & -\sin\theta_\mathbf{k} e^{-i\varphi_\mathbf{k}} \\
        \sin\theta_\mathbf{k} e^{i\varphi_\mathbf{k}}    &  \cos\theta_\mathbf{k}
        \end{array}
\right).
\end{equation}
Here $\theta_\mathbf{k}\in[0,\pi/2]$, $\varphi_\mathbf{k}\in[0,2\pi)$. The definitions of $\theta_\mathbf{k}$ and $\varphi_\mathbf{k}$ are given by
\begin{equation}
\cos(2\theta_\mathbf{k}) = \left\{
    \begin{aligned}
        & 1, & & \mathbf{k}=0 \\
        & k_z/|\mathbf{k}|, & &\text{otherwise}
    \end{aligned}
\right. ,
\end{equation}
 and
\begin{equation}
    e^{i\varphi_\mathbf{k}} = \left\{
    \begin{aligned}
        & 1, \ \ \ k_x=k_y=0 \\
        & (k_x+ik_y)/|k_x+ik_y|, \ \ \ \text{otherwise}
    \end{aligned}
\right. .
\end{equation}
After the above transformation of  local spin rotation in the k-space, $\hat{H}_0=\sum_{\mathbf{k},s=\pm} \varepsilon_{\mathbf{k},s}^\mathbf{b} \hat{n}_{\mathbf{k},s}$ is diagonalized with $\hat{n}_{\mathbf{k},s}=a_{\mathbf{k},s}^\dag a_{\mathbf{k},s}$ and
\begin{equation}
    \varepsilon_{\mathbf{k},s}^\mathbf{b} = \varepsilon_\mathbf{k}+s\alpha|\mathbf{k+b}|. \label{sp}
\end{equation}
From the above equation, it is clearly seen that for constant or weak $\mathbf{k}$-dependent $\varepsilon_\mathbf{k}$
$\varepsilon_{\mathbf{k-b},s}^\mathbf{b} \approx \varepsilon_{\mathbf{-k-b},s}^\mathbf{b}$, so that the intra-branch
pairing $(\mathbf{k-b},s;\mathbf{-k-b},s)$ with the total momentum $-2\mathbf{b}$ is favored in the presence of the effective Zeeman field and
the spin-orbit interaction.
The idea is schematically illustrated in Fig.~\ref{demo}. For general $\varepsilon_\mathbf{k}$, the nesting vector $\mathbf{q}$, satisfying
$\varepsilon_{\mathbf{k+q},s}^\mathbf{b} \approx \varepsilon_{\mathbf{-k+q},s}^\mathbf{b}$, is not equal to $-\mathbf{b}$ but
still proportional to the Zeeman field \cite{note1}.
Thus with the nesting vector $\mathbf{q}$, the reduced BCS Hamiltonian
is singled out from a group of candidates, which is related to the FFLO state~\cite{FFLO,Dukelsky,note1},
\begin{equation}
    \hat{H}_\text{int}(\mathbf{q})= V_{0}\sum_{\mathbf{k},\mathbf{k}^\prime}
        c_{\mathbf{k+q}\uparrow}^{\dag} c_{\mathbf{-k+q}\downarrow}^{\dag}
        c_{\mathbf{-k^\prime+q}\downarrow}c_{\mathbf{k^\prime+q}\uparrow}.
    \label{Hint}
\end{equation}
The above equation can be rewritten in terms of $a_\mathbf{k,s}^\dag$ and $a_\mathbf{k,s}$ by applying the unitary transformation
and accordingly the pairing Hamiltonian becomes
\begin{equation}
    \hat{H}_P = \hat{H}_0 + V_0 \sum_{\mathbf{k}s,\mathbf{k}^\prime s^\prime}  e^{-is\varphi_\mathbf{k}+is^\prime\varphi_{\mathbf{k}^\prime}}
                A_{\mathbf{k},s}^\dagger(\mathbf{q}) A_{\mathbf{k}^\prime,s^\prime}(\mathbf{q}),
    \label{pairingHam}
\end{equation}
where $\hat{H}_0$ is re-expressed as
\begin{equation}
    \hat{H}_0 = \sum_{\mathbf{k},s=0,\pm} \xi_{\mathbf{k},s}^\mathbf{b,q} \hat{n}_{\mathbf{k},s}^\mathbf{q}
                                                + \eta_{\mathbf{k},s}^\mathbf{b,q} \hat{m}_{\mathbf{k},s}^\mathbf{q}
\end{equation}
with
$$
    \xi(\eta)_{\mathbf{k},s}^{\mathbf{b},\mathbf{q}} = \left\{
        \begin{aligned}
            & \frac{\varepsilon_{\mathbf{k+q},s}^\mathbf{b} +(-) \varepsilon_{\mathbf{-k+q},s}^\mathbf{b}}{2},& \ &\mathbf{k}\in K_1, s=\pm \\
            & \frac{\varepsilon_{\mathbf{k+q},+}^\mathbf{b} +(-) \varepsilon_{\mathbf{-k+q},-}^\mathbf{b}}{2},& \  & \mathbf{k}\in K_0, s=0 \\
            & 0,& \ &\text{otherwise}
        \end{aligned}
    \right.
$$
and
$$
    \hat{n}(\hat{m})_{\mathbf{k},s}^\mathbf{q} = \left\{
        \begin{aligned}
            & \hat{n}_{\mathbf{k+q},s} +(-) \hat{n}_{\mathbf{-k+q},s},&\ &\mathbf{k}\in K_1, s=\pm \\
            & \hat{n}_{\mathbf{k+q},+} +(-) \hat{n}_{\mathbf{-k+q},-},&\ &\mathbf{k}\in K_0, s=0 \\
            & 0,&\ &\text{otherwise}
        \end{aligned}
    \right. .
$$
In the above formulas, $K_0\equiv\{\mathbf{k}|(k_x,k_y,k_z)=0\}$ and $K_1\equiv \{\mathbf{k}|k_x=k_y=0, k_z>0 \} \cup \{\mathbf{k}|k_x = 0,k_y>0 \} \cup \{\mathbf{k}|k_x > 0 \}$. The pair operator in Eq.~(\ref{pairingHam}) is defined as,
\begin{equation}
A_{\mathbf{k},s}^\dagger(\mathbf{q}) = \left\{
    \begin{aligned}
        & a_{\mathbf{k+q},s}^\dagger a_{\mathbf{-k+q},s}^\dagger,&\ \ \ & \mathbf{k}\in K_1\text{ and } s=\pm \\
        & a_{\mathbf{k+q},+}^\dagger a_{\mathbf{-k+q},-}^\dagger,&\ \ \ &\mathbf{k}\in K_0\text{ and } s=0 \\
        & 0,&\ \ \ &\text{otherwise}
    \end{aligned}
\right. .
\end{equation}
In the derivation of Eq.~(\ref{pairingHam}), we have used the relations $\cos\theta_{\mathbf{-k}}=\sin\theta_\mathbf{k}$ and $e^{i\varphi_\mathbf{-k}}=-e^{i\varphi_\mathbf{k}}$
for $(k_x,k_y)\neq 0$.
Note that the definitions of the above pair operators are slightly different from those in the 2D case~\cite{JLiu}, which plays a crucial role in solving the 3D model exactly.

The operators involved in the pairing Hamiltonian satisfy the following commutation relations
\begin{equation}
\begin{aligned}
   & [\hat{n}_{\mathbf{k},s}^\mathbf{q}, A_{\mathbf{k}^\prime,s^\prime}^\dagger(\mathbf{q})]
        = 2 \delta_{\mathbf{k}s,\mathbf{k}^\prime s^\prime} A_{\mathbf{k},s}^\dagger(\mathbf{q}),\\
    & [A_{\mathbf{k},s}(\mathbf{q}),A_{\mathbf{k}^\prime,s^\prime}^\dagger(\mathbf{q})]
        = \delta_{\mathbf{k}s,\mathbf{k}^\prime s^\prime} ( 1 - \hat{n}_{\mathbf{k},s}^\mathbf{q} ), \\
    & [\hat{m}_{\mathbf{k},s}^\mathbf{q}, \hat{n}_{\mathbf{k}^\prime,s^\prime}^\mathbf{q}] =
        [\hat{m}_{\mathbf{k},s}^\mathbf{q}, A_{\mathbf{k}^\prime,s^\prime}(\mathbf{q})] =  0,
\end{aligned}
\label{commfermion}
\end{equation}
and therefore the commutator algebra is closed.
Using a scenario similar to that employed in Ref.~\cite{JLiu}, the eigenstates of the Hamiltonian Eq.~(\ref{pairingHam}) are found to be
\begin{equation}
    |n,S_{+},S_{-}\rangle= \prod_{\mathbf{k}_{i}\in
    S_{+}}a_{\mathbf{k}_{i}+\mathbf{q},+}^{\dag} \prod_{\mathbf{k}_{j}\in
    S_{-}}a_{\mathbf{k}_{j}+\mathbf{q},-}^{\dag}
    \prod_{\nu=1}^{n}B_{\nu}^{\dag}|0\rangle \label{eigenFFLO}
\end{equation}
with
\begin{equation}
B_{\nu}^{\dag} = \sum_{
                        \begin{subarray}{c}
                            s=0,\pm; \\
                            \mathbf{k}\in P_s
                        \end{subarray}
                    }
    \frac{e^{-i s\varphi_\mathbf{k}}A_{\mathbf{k},s}^\dag(\mathbf{q})}{2\xi_{\mathbf{k},s}^\mathbf{b,q}-E_{\nu}},
\end{equation}
where the parameters $E_\nu$ ($\nu=1,2,\ldots,n$) are solutions of the $n$-coupled Richardson's equations \cite{richardson64}
\begin{equation}
\begin{aligned}
    1+\sum_{
            \begin{subarray}{c}
                s=0,\pm; \\
                \mathbf{k}\in P_s
            \end{subarray}
        }    \frac{V_{0}}{2\xi_{\mathbf{k},s}^\mathbf{b,q}-E_\nu}
        - \sum_{\mu\neq\nu}^{n}\frac{2V_{0}}{E_\mu-E_\nu}=0.
\end{aligned}
\label{req3D}
\end{equation}
Here $S_\pm$ denotes the set of singly occupied levels (namely
blocked levels) of the $\pm$ branch with cardinality $m_\pm$, while
$P_\pm$ and $P_0$ the sets of levels with the blocked ones excluded. The state
vector defined in Eq.~(\ref{eigenFFLO}) describes the eigenstates of
$N_{e}=m_{+}+m_{-}+2n$ fermions with $n$ as the number of pairs.
The corresponding energy is
\begin{equation}
\begin{aligned}
    E(n,S_+,S_-)
            = & \sum_{\mathbf{k} \in S_+ } \varepsilon_{\mathbf{k+q},+}^\mathbf{b} + \sum_{\mathbf{k} \in S_- } \varepsilon_{\mathbf{k+q},-}^\mathbf{b}
            + \sum_{\nu=1}^n E_\nu.
\end{aligned}
\label{totalE}
\end{equation}

For the special case of $\mathbf{B}=0$, it is straightforward to conclude that the SO-coupled 3D pairing fermion Hamiltonian is
exactly solvable, as in the 2D case \cite{JLiu}.
Several remarks on the present 3D case can be made.
(i) We calculate the gap function (as well as the pair-pair correlation function) as we did in the 2D case, and
find that $\Delta_{\mathbf{k},s}=e^{-is\varphi_\mathbf{k}} \Delta_{\mathbf{k},s}^0$, where  $\Delta_{\mathbf{k},s}^0$ is always real and just the usual $s$-wave one.
Therefore the pairing symmetry of the gap function $\Delta_{\mathbf{k},s}$ is of $p_x+ip_y$-wave.
(ii) We can indeed show from Eq.~(\ref{eigenFFLO}) that the pairing ground state has the time-reversal symmetry, as expected by the MFT.
(iii) The results can be readily generalized to the cases with anisotropic spin-orbit interactions
$(\alpha_x k_x, \alpha_y k_y, \alpha_z k_z)\cdot{\bm\sigma}$ by a scale transformation of $\mathbf{k}$.
(iv) If $k_z=0$, we are able to
recover those results obtained in our previous 2D work~\cite{JLiu}.
(v) It was indicated by Gaudin that the Richardson's equations (\ref{req3D})
can recover the BCS gap and number equations in the continuum limit under the single-arc assumption~\cite{gaudin,roman}.

Most interestingly, when $\varepsilon_\mathbf{k}$ is a $\mathbf{k}$-independent constant, the Hamiltonian for $\mathbf{B}\neq 0$ can be related to that for $\mathbf{B}=0$ by the following simple mappings:
\begin{equation}
\begin{aligned}
    & d_\mathbf{k,s} = a_{\mathbf{k-b},s}, \\
    & \tilde{A}_{\mathbf{k},s}^\dag = A_{\mathbf{k},s}^\dag(\mathbf{b}) = \left\{
    \begin{aligned}
        & d_{\mathbf{k},s}^\dagger d_{\mathbf{-k},s}^\dagger,&\ \ \ & \mathbf{k}\in K_1\text{ and } s=\pm \\
        & d_{\mathbf{k},+}^\dagger d_{\mathbf{-k},-}^\dagger,&\ \ \ &\mathbf{k}\in K_0\text{ and } s=0 \\
        & 0,&\ \ \ &\text{otherwise}
    \end{aligned}
    \right.
\end{aligned}
\nonumber
\end{equation}
for the single-particle and pair operators, respectively.
From this exact mapping, we have for the ground state with $n$ pairs and no unpaired fermions,
\begin{equation}
\begin{aligned}
&    E_\nu(\mathbf{B}) = E_\nu(\mathbf{B}=0),\ \nu=1,2,\cdots,n, \\
&    |n\rangle_\text{FF}(\mathbf{B}) = \text{T}_{\mathbf{-b}} |n\rangle(\mathbf{B}=0) , \\
&    E(n,\mathbf{B}) = E(n,\mathbf{B}=0),
\end{aligned}
\end{equation}
where $\text{T}_{\mathbf{-b}}$ means the translation of each fermion pair by $-2\mathbf{b}$.
The peculiarity of this type of FF state is that the center-of-mass momentum of the Cooper pair is $\mathbf{Q}=-2\mathbf{b}$ with its order parameter being proportional to  $e^{i\mathbf{Q}\cdot\mathbf{R}}$. This
drifting effect of the Zeeman field on the superfluid of SO-coupled (or Dirac-type) fermions is similar to that of the vector potential of the magnetic field
on a superconductor, but without having to increase its kinetic energy. And due to the nature of Dirac fermions, the mechanism of the above FF state is definitely different from that accounting for conventional FF states.
For the isotropic SO-interaction, $\mathbf{Q}$ align in the same direction of the Zeeman field, while for the anisotropic
case, $\mathbf{Q}$ is along the direction of the scaled Zeeman field.

\begin{figure}[ht]
\includegraphics[width=6cm]{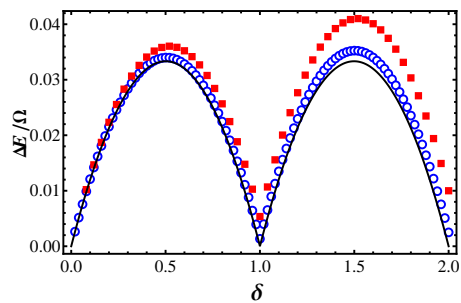}
\caption{The scaled condensation energy $\Delta E/\Omega$ in units of $\alpha k_\text{F}$ as a function of the filling factor $\delta=N/\Omega$ with
$\Omega=50$ (red squares) and $\Omega=200$ (blue circles) calculated from exact solutions. The scaled interaction is fixed with $V_0\Omega=0.5$. The solid black curve is the results from the MFT. }
\label{scaling}
\end{figure}

As a simple example, we consider a special case where all $N$ fermions are on the Fermi
surface $|\mathbf{k+b}|=k_\text{F}$, which may be viewed as an approximation for the considered system
as many physical phenomena
may be mainly related to the electrons near the Fermi surface. The degeneracy of the Fermi level is
assumed to be $\Omega$. In the presence of the SO-interaction,
the energy level is split into two branches $\varepsilon_{\mathbf{k},\pm}^\mathbf{b} = \varepsilon_F \pm \alpha k_\text{F}$.
The symbols in Fig.~\ref{scaling} denote the scaled condensation energy $\Delta E/\Omega$ as a function of the fermion density for two values of $\Omega$ and
the scaled interaction fixed, calculated from
Eqs.~(\ref{req3D}) and (\ref{totalE}) numerically. These exact results of $\Delta E/\Omega$ are always higher than those obtained from the MFT and approach to them in the large $\Omega$ limit, as shown in the figure. Moreover, some results are beyond the expectation of the MFT. We find that the exact condensation energy $\Delta E$ is always finite at half filling ($\delta=1$), indicating a superconducting state in contrast to the normal-state (for $V_0\Omega < 2$) %quantum critical point
predicted by the MFT. The exact data is obviously asymmetric with respect to $\delta=1$, while the MFT curve displays a clear particle-hole symmetry.
These differences between the exact solution and the MFT are more notable for small $N$ cases and therefore more significant for cold-atom systems.

% boson case
At this stage, we generalize the drifting effect of the Zeeman field to the SO-coupled spinor Bose-Einstein condensates,
considering that the SO coupling has been realized in the ultracold bosonic gas lately~\cite{lin2011}. The Hamiltonian of the interacting Boson gas with SO and Zeeman couplings is taken to be the same as Eq.~(\ref{ham}) except that $c_{\mathbf{k}\uparrow(\downarrow)}$'s are now annihilation operators of bosons and $V_0 \geqslant 0$ indicating  a repulsive interaction between bosons.
The single-particle spectrum consists of two branches as given by Eq.~(\ref{sp}) with $\varepsilon_\mathbf{k}=\mathbf{k}^2/2m$.
According to the dispersion $\varepsilon_{\mathbf{k},\pm}$, one can readily find that the ground state of the non-interacting ($V_0=0$) ideal boson gas corresponds to the condensation of all bosons
in the lowest-lying single-particle state. This energy minimum locates at $\mathbf{k}_0 = m \alpha \hat{\mathbf{B}}$ in the lower branch ($s=-1$) with   $\hat{\mathbf{B}}$ denotes the direction of the Zeeman field. Therefore, the role of the Zeeman field is to lock the direction of the momentum of the condensate.
Considering the condensation of bosons, we can truncate the interaction Hamiltonian and obtain the exactly solvable pairing Hamiltonian for 2D bosons \cite{lieb,richardson68,Dukelsky01}, which reads
\begin{equation}
    H = \sum_{\mathbf{k},s=\pm} \varepsilon_{\mathbf{k},s} a_{\mathbf{k},s}^\dagger a_{\mathbf{k},s} + V_0 \sum_{\mathbf{k},\mathbf{k}^\prime}
                A_{\mathbf{k}}^\dagger(\mathbf{k}_0) A_{\mathbf{k}^\prime}(\mathbf{k}_0),
    \label{hamb}
\end{equation}
with $s$-wave inter-branch bosonic pairing
$$A_{\mathbf{k}}^\dagger(\mathbf{k}_0) = a_{\mathbf{k+k}_0,+}^\dagger a_{\mathbf{-k+k}_0,-}^\dagger,$$
which is distinct from the chiral $p$-wave intra-branch pairing
of the fermionic case, due to different commutation relations.
This pairing Hamiltonian takes into account the effect of the interaction on the depletion of BEC owning to the inter-branch pair scattering process of
bosons from the lowest-lying $\mathbf{k}_0$ state to the excited states. Note that the Hamiltonian is still exact solvable even after including the pair scattering. According to Richardson's ansatz~\cite{richardson68}, the eigenstates take the form as
\begin{equation}
    |n,\{m_{\mathbf{k},s}\}\rangle= \prod_{\nu=1}^{n} B_{\nu}^{\dag} \prod_{\mathbf{k},s} (a_{\mathbf{k}+\mathbf{k}_0,s}^\dagger)^{m_{\mathbf{k},s}} |0\rangle, \label{eigen}
\end{equation}
where
\begin{equation}
B_\nu^\dagger = \sum_{\mathbf{k}} \frac{ A_{\mathbf{k}}^\dagger(\mathbf{k}_0)}{\varepsilon_{\mathbf{k+k}_0,+}^\mathbf{b}+\varepsilon_{\mathbf{-k+k}_0,-}^\mathbf{b}-E_\nu}
\end{equation}
and $m_{\mathbf{k},s}$ denotes the number of unpaired bosons. Here $E_\nu$'s obey the following Richardson's
equation,
\begin{equation}
\begin{aligned}
    1+V_{0}\sum_{\mathbf{k}}
        \frac{1+m_{\mathbf{k},+}+m_{\mathbf{-k},-}}{\varepsilon_{\mathbf{k+k}_0,+}^\mathbf{b}+\varepsilon_{\mathbf{-k+k}_0,-}^\mathbf{b}-E_\nu}
        +\sum_{\mu\neq\nu}^{n}\frac{2V_{0}}{E_\mu-E_\nu}=0.
\end{aligned}
\label{richb}
\end{equation}
Despite the seemingly analogy of the pairing Hamiltonian~(\ref{pairingHam}) and (\ref{hamb}) as well as the resulting Richardson's equations (\ref{req3D}) and (\ref{richb}), some significant differences should be noted between the fermion and boson systems. The summation over $\mathbf{k}$ in the second term of the Richardson's equation excludes (includes) the unpaired states and the sign preceding the third term is $-$ ($+$)  for the fermion (boson) case. This difference originates from the important
sign change in the commutation relations, namely $[A_\mathbf{k}(\mathbf{k}_0),A_{\mathbf{k}'}^\dagger(\mathbf{k}_0)]=\delta_{\mathbf{k},\mathbf{k}'}[1+\sum_{s=\pm}\hat{n}_{s\mathbf{k+k}_0,s}]$ for the boson case, in contrast to Eq.~(\ref{commfermion}) for the fermion case.
As for the similar current-carrying ground state, the underlying physics for both cases is also quite different.
For the fermion case, the fermiology governs the ground state (at least in the weak-coupling limit) and
the nesting of the shifted and/or deformed Fermi surface by the Zeeman field favors the Cooper pairing with nonzero center-of-mass momentum; while for the boson case, the physics is dominated by the tendency of occupation of all bosons on the lowest-lying state whose position is determined by the Zeeman field.

To conclude, we have studied the BCS-type pairing model of Zeeman-coupled Dirac-type fermions in three dimensions and found that the model is exactly solvable.
%In the absence of the Zeeman field, our results are the generalization from 2D to 3D SO-coupled superfluidity/superconductors. While for the nonzero Zeeman field,
In particular, we have revealed rigorously an unconventional FFLO ground state with the translational momentum of the Cooper pairs oriented in the direction of the Zeeman field. An analogue ground state has also been studied for the SO- and Zeeman-coupled Bose-Einstein condensate.
% where the direction of the drifting momentum is also found
%to align in that of the Zeeman field.

We would like to thank  Y.~C.~He, Y. Chen, T. Mao, B. Wu, and J. R. Shi for
helpful discussions. ZDW also thanks the ICQM at Peking University for hospitality
to his visit there, during which this work was finalized.
This work was supported by the RGC of Hong Kong
(Nos. HKU7044/08P, HKU7055/09P, and HKU7058/11P) and a CRF of Hong Kong. The Natural Science Foundation
of China No.~10674179 and the SKPBR of China (No.~2011CB922104).


\begin{thebibliography}{90}


\bibitem{Read} N.~Read and D.~Green, Phys.~Rev.~B \textbf{61}, 10267 (2000).
\bibitem{Fu} L.~Fu and C.~L.~Kane, Phys.~Rev.~Lett.~\textbf{100}, 096407(2008).
\bibitem{Sau} J.~D.~Sau, R.~M.~Lutchyn, S.~Tewari  and S.~Das Sarma, Phys.~Rev.~Lett.~\textbf{104}, 040502 (2010).

\bibitem{JLiu} J. Liu, Q. Han, L. B. Shao and Z. D. Wang, \prl~\textbf{107}, 026405 (2011).

\bibitem{Hor} Y. S. Hor {\it et al.}, \prl~\textbf{104}, 057001 (2010).
\bibitem{Wray} L.A. Wray {\it et al.}, Nature Phys.~\textbf{6}, 855 (2010).
\bibitem{Kriener} M. Kriener, K. Segawa, Z. Ren, S. Sasaki, and Yoichi Ando, \prl~\textbf{106}, 127004 (2011).

\bibitem{zhu1}S.~L.~Zhu, D.~W.~Zhang and Z.~D.~Wang, Phys.~Rev.~Lett.~\textbf{102}, 210403 (2009).

\bibitem{zhu2}S.~L.~Zhu, L.~B.~Shao, Z.~D.~Wang and L.~M.~Duan, Phys.~Rev.~Lett.~\textbf{106}, 100404 (2011).

\bibitem{lin2011}  Y.~J.~Lin, K. Jimenez-Carcia, and I. B. Spielman, Nature \textbf{471}, 83 (2011).

\bibitem{zhu3} M.~Yang and S.~L.~Zhu, Phys.~Rev.~A \textbf{82}, 064102 (2010).

\bibitem{FFLO} P. Fulde and R. A. Ferrell, Phys. Rev. \textbf{135}, A550 (1964); A. I. Larkin and Yu. N. Ovchinnikov, Sov. Phys. JETP
\textbf{20}, 762 (1965).

\bibitem{Yokoyama} T.~Yokoyama, Y.~Tanaka, and N.~Nagaosa, Phys.~Rev.~B \textbf{81}, 121401(R) (2010).

\bibitem{Liu} X.-J. Liu, M. F. Borunda, X. Liu, and J. Sinova, \prl~\textbf{102}, 046402 (2009).

\bibitem{note1} Q. Han {\it et al.}, arXiv:1104.0614 (2011).

\bibitem{Dukelsky} J.~Dukelsky, G.~Ortiz, S.M.A.~Rombouts, and K.~Van Houcke, \prl~\textbf{96}, 180404 (2006).

\bibitem{richardson64} R.~W.~Richardson and N.~Shermann, Nucl.~Phys.~\textbf{52}, 221 (1964).
%Note that the original formula given
%by Richardson and Shermann is not valid when $n > \tilde{\Omega}$ and we generalize it as shown in the text.

\bibitem{gaudin} M.~Gaudin, Travaux de Michel Gaudin, Mod\`{e}les Exactament R\`{e}solus, Les \`{E}ditions de Physique, France, (1995).


\bibitem{roman} J.~M.~Roman, G.~Sierra and J.~Dukelsky, Nucl.~Phys.~B.~\textbf{634}, 483 (2002).


\bibitem{lieb} E.~H.~Lieb, {\it Lectures in Theoretical Physics} (The University of Colorado Press, Boulder, Colo. 1965), Vol. VIIc, p. 175.

\bibitem{richardson68} R.~W.~Richardson, J.~Math.~Phys. \textbf{9}, 1327 (1968).

\bibitem{Dukelsky01} J.~Dukelsky and P.~Schuck, \prl~\textbf{86}, 4207 (2001).

\end{thebibliography}
\end{document}